\documentclass[aps,prb,twocolumn,superscriptaddress,longbibliography,floatfix,nofootinbib]{revtex4-2}

\usepackage{graphicx}
\graphicspath{{figures/}}
\usepackage{amsmath,amssymb}
\usepackage{bm}
\usepackage{booktabs}
\usepackage{mathtools}
\usepackage{array}
\usepackage[colorlinks=true,linkcolor=blue,citecolor=blue,urlcolor=blue]{hyperref}

\newcommand{\sx}{\hat{\tau}_x}
\newcommand{\sy}{\hat{\tau}_y}
\newcommand{\sz}{\hat{\tau}_z}
\newcommand{\sI}{\hat{\tau}_0}
\newcommand{\tauh}[1]{\hat{\tau}_{#1}}
\newcommand{\tr}{\mathrm{Tr}}

\newcommand{\ket}[1]{|#1\rangle}

\begin{document}

\title{Dissipation-driven boundary localization in higher-order topological insulators}

\author{Xue-Ping Ren}
\thanks{These authors contributed equally to this work.}
\affiliation{Center for Advanced Quantum Studies, School of Physics and Astronomy, Beijing Normal University, Beijing 100875, China}
\affiliation{Key Laboratory of Multiscale Spin Physics (Ministry of Education), Beijing Normal University, Beijing 100875, China}

\author{Xiao-Ran Wang}
\thanks{These authors contributed equally to this work.}
\affiliation{College of Physics and Hebei Advanced Thin Film Laboratory,Hebei Normal University, Shijiazhuang 050024, China}

\author{Xin-Ran Ma}
\affiliation{Center for Advanced Quantum Studies, School of Physics and Astronomy, Beijing Normal University, Beijing 100875, China}
\affiliation{Key Laboratory of Multiscale Spin Physics (Ministry of Education), Beijing Normal University, Beijing 100875, China}

\author{Xi Hu}
\affiliation{Center for Advanced Quantum Studies, School of Physics and Astronomy, Beijing Normal University, Beijing 100875, China}
\affiliation{Key Laboratory of Multiscale Spin Physics (Ministry of Education), Beijing Normal University, Beijing 100875, China}

\author{Su-Peng Kou}
\email{spkou@bnu.edu.cn}
\affiliation{Center for Advanced Quantum Studies, School of Physics and Astronomy, Beijing Normal University, Beijing 100875, China}
\affiliation{Key Laboratory of Multiscale Spin Physics (Ministry of Education), Beijing Normal University, Beijing 100875, China}

\date{\today}

\begin{abstract}
We study dissipation-driven boundary localization in a Bernevig--Hughes--Zhang-type second-order topological insulator by introducing inhomogeneous, spin-dependent boundary dissipation. After projection onto the edge subspace, the dominant component of the boundary dissipation lies in the same Pauli channel as the kinetic term. This channel gives both counter-propagating edge modes the same real transfer exponent. The two modes therefore accumulate at the same dissipative domain wall. For the corner states, the dissipative envelope competes with Jackiw--Rebbi localization. Increasing dissipation moves their weight from the geometric corners to the dissipative domain wall. Propagating edge states also localize at this wall. Numerical calculations confirm the qualitative predictions of the edge theory. A fixed-energy Feshbach reduction captures finite-size effects. Inhomogeneous boundary dissipation thus controls boundary-state localization without changing the bulk topology.
\end{abstract}
\maketitle

\section{Introduction}\label{sec:intro}

\enlargethispage{0.95\baselineskip}
Non-Hermitian Hamiltonians describe open systems with gain or loss and can support complex spectra, exceptional points, and line- or point-gap topology~\cite{Bender1998,Heiss2012,Rotter2009,ElGanainy2018,Miri2019,Ashida2020,Bergholtz2021,Gong2018,Kawabata2019symmetry}. A key phenomenon is the non-Hermitian skin effect (NHSE), in which nonreciprocity drives an extensive number of bulk states toward a boundary~\cite{HatanoNelson1996}. This accumulation invalidates the conventional Bloch bulk--boundary correspondence and motivates non-Bloch band theory~\cite{Yao2018edge,Yokomizo2019,Okuma2020,Zhang2020correspondence,Yang2020}. Most studies introduce non-Hermiticity into the bulk Hamiltonian, both theoretically and experimentally~\cite{Lee2016,Leykam2017,Shen2018,Kunst2018,MartinezAlvarez2018,LeeThomale2019,Song2019,Longhi2019,Weidemann2020,Xiao2020,Ghatak2020,Helbig2020,Hofmann2020,Li2020}. In contrast, non-Hermiticity can be confined to an open boundary, where it can generate boundary point-gap topology~\cite{Schindler2023NHboundary}.  Inhomogeneous boundary dissipation causes  the non-Hermitian chiral skin effect tht a single chiral edge mode to accumulate at a dissipative domain wall~\cite{Ma2024}. Structured loss can further select the boundary segment or geometric corner where this edge-state accumulation occurs~\cite{Liu2024chiral}. These studies establish dissipative-domain-wall localization for a single chiral boundary channel.

This mechanism has also been extended to a gapless helical edge, under staggered loss, the two pseudospin sectors accumulate at opposite dissipative domain walls in a phononic Kane--Mele system~\cite{Wu2024}. Higher-order topological insulators (HOTIs) provide a natural setting in which this decoupled picture no longer holds~\cite{Benalcazar2017science,Benalcazar2017prb,Schindler2018,Song2017,Langbehn2017}. In a two-dimensional second-order topological insulator (SOTI), a Dirac mass couples the two counter-propagating components and opens an edge gap. Adjacent gapped edges carry opposite Dirac masses, whose sign changes at the corners bind Jackiw--Rebbi zero modes when chiral symmetry is present~\cite{Jackiw1976,Khalaf2018,Geier2018,Yan2019}. Such corner states have been observed in topolectrical circuits and microwave structures~\cite{Imhof2018,Peterson2018}. Non-Hermitian higher-order systems can also exhibit higher-order skin localization and gain/loss-induced corner states~\cite{Lee2019hybrid,Kawabata2020higher,Zhang2022universal,Liu2019,Luo2019}, and their bulk--boundary correspondence and second-order skin effects have been studied~\cite{Edvardsson2019,Ezawa2019NHHOTI,Okugawa2020,Zou2021}. These studies mainly concern modified higher-order topology, non-Hermiticity-induced corner states, or higher-order skin accumulation. It remains unclear whether the two components of a mass-coupled helical edge still accumulate at opposite dissipative domain walls or instead move toward the same wall. It is also unclear whether boundary dissipation can overcome mass-domain-wall confinement and redistribute the pre-existing Jackiw--Rebbi corner states without modifying the periodic bulk Hamiltonian.

In this paper, we address these questions in a BHZ-type SOTI with inhomogeneous, spin-dependent boundary dissipation. The research results indicate that the chiral-skin mechanism survives the mass coupling but follows a modified localization rule. After projection onto the edge subspace, the dominant part of the boundary dissipation lies in the same Pauli channel as the kinetic term. Between the two counter-propagating components, the propagation velocity and the projected dissipation reverse sign together, producing the same real transfer exponent. Both components therefore accumulate at the same dissipative domain wall rather than at opposite walls. The resulting common dissipative envelope also acts on the corner states, where it competes with Jackiw--Rebbi localization. As the dissipation increases, the corner-state weight shifts from the geometric corners toward the same dissipative domain wall. Numerical calculations confirm these qualitative predictions. A fixed-energy Feshbach reduction captures the finite-size deviations from the ideal straight-edge theory. Inhomogeneous boundary dissipation therefore provides a boundary-only way to control the localization of existing SOTI boundary states without changing the bulk topology.

The rest of the paper is organized as follows. Section~\ref{sec:model} introduces the lattice model and its effective edge Hamiltonian. Section~\ref{sec:edge} develops the boundary transfer theory and analyzes the localization of propagating and corner states in different dissipation channels. Section~\ref{sec:numerics} presents the numerical results and compares them with the edge theory and the fixed-energy Feshbach reduction. Section~\ref{sec:conclusion} summarizes the main findings.

\section{Model}\label{sec:model}

We use a BHZ-type lattice model with a time-reversal-breaking mass term that gaps the helical edges:
\begin{equation}\label{eq:bulk}
\begin{aligned}
&H_0(\bm{k})=d_0(\bm{k})\,\sigma_z s_0+A\sin k_x\,\sigma_x s_z+A\sin k_y\,\sigma_y s_0\\
&+\Lambda(\cos k_x-\cos k_y)\,\sigma_x s_x,
\end{aligned}
\end{equation}
where
\begin{equation}
d_0(\bm{k})=M-2B(2-\cos k_x-\cos k_y).
\end{equation}
The Pauli matrices $\sigma_i$ and $s_i$ act on orbital and spin degrees of freedom.  At $\Lambda=0$ and $0<M<4B$, the model is a quantum spin Hall insulator with time-reversal symmetry $\mathcal T=is_y\mathcal K$ and a nontrivial $\mathbb Z_2$ topological invariant~\cite{Bernevig2006,Konig2007}. The $\Lambda$ term breaks time-reversal symmetry and opens a gap in the helical edge spectrum.  Nevertheless, the Hamiltonian retains a unitary chiral symmetry:
\begin{equation}\label{eq:chiralsym}
\mathcal S H_0(\bm k)\mathcal S^{-1}=-H_0(\bm k),\qquad \mathcal S=\sigma_xs_y,
\end{equation}
and a fourfold rotation
\begin{equation}\label{eq:c4sym}
\begin{aligned}
&\mathcal R_4=s_z\exp\!\left(-\frac{i\pi}{4}\sigma_zs_z\right),\\
&\mathcal R_4^4=-1,\\
&\mathcal R_4H_0(k_x,k_y)\mathcal R_4^{-1}=H_0(-k_y,k_x).
\end{aligned}
\end{equation}
Under $C_4$, both $\cos k_x-\cos k_y$ and $\sigma_xs_x$ change sign, so their product is invariant.  After projection onto the edge states, this term gives opposite Dirac masses on the edges parallel to the $x$ and $y$ axes.  Chiral symmetry keeps an isolated Jackiw--Rebbi mode at zero energy, while $C_4$ symmetry makes the edge mass alternate in sign between neighboring edges of an ideal square.

The non-Hermitian perturbation is applied only to boundary sites,
\begin{equation}\label{eq:boundaryNH}
\delta H=i\sum_{\bm r\in\partial\Omega}\gamma(\bm r)\, c_{\bm r}^\dagger s_z c_{\bm r}.
\end{equation}
Because this perturbation is introduced only after the boundary is opened, it leaves the periodic-boundary Bloch Hamiltonian $H_0(\bm k)$ unchanged.  Here $\gamma(\bm r)$ is the real bare lattice gain/loss coefficient and is piecewise constant along the boundary, thereby implementing the inhomogeneous boundary dissipation studied below.

Projecting Eqs.~\eqref{eq:bulk} and~\eqref{eq:boundaryNH} onto the low-energy boundary subspace gives the effective one-dimensional Hamiltonian
\begin{equation}\label{eq:hedge}
H_{\rm edge}=-iA_{\rm e}\partial_\ell\,\sz+m_{\rm e}(\ell)\,\sy-i\gamma_{\rm e}(\ell)\,\sz,
\end{equation}
where $\ell$ is the coordinate along the closed boundary, with the counterclockwise direction defined as positive.  The Pauli matrices $\sx,\sy,\sz$ act on the two low-energy edge channels, which propagate in opposite directions.  The projected parameters $A_{\rm e}$ and $m_{\rm e}(\ell)$ are the effective edge velocity and effective edge mass, and $\gamma_{\rm e}(\ell)$ is the effective dissipation strength in the kinetic channel.  For a finite square lattice, mass profile is
\begin{equation}\label{eq:mass}
m_{\rm e}(\ell)\simeq m_0\eta(\ell),\qquad m_0=\frac{\Lambda M}{2B},
\end{equation}
Here, $m_0$ is the leading-order continuum estimate of the mass magnitude.  The function $\eta(\ell)$ changes sign between adjacent edges.  Each sign change forms a mass domain wall (MDW) at a corner.

In the counterclockwise gauge, the sign of the dissipation term in Eq.~\eqref{eq:hedge} is determined to be negative,  and the derivation can be found in Appendix~\ref{app:projection}.        

Eq.~\eqref{eq:hedge} describes only low-energy boundary states.  It can be used quantitatively only when
\begin{equation}\label{eq:validity}
|\gamma_{\rm e}|,\ |m_{\rm e}|,\ |E|\ll E_{\rm bulk},
\end{equation}
where $E_{\rm bulk}$ is the energy from zero to the nearest bulk continuum.

\section{Boundary localization from the edge theory}\label{sec:edge}

To describe localization on the closed boundary, we separate the projected dissipation into its boundary average and spatially varying part,
\begin{equation}\label{eq:globalgamma}
\bar\gamma_{\rm e}=\frac{1}{\mathcal L}\oint\gamma_{\rm e}(\ell)d\ell,\qquad \gamma_{{\rm g},{\rm e}}(\ell)=\gamma_{\rm e}(\ell)-\bar\gamma_{\rm e}.
\end{equation}
Here, $\mathcal L$ is the total boundary length.  The average $\bar\gamma_{\rm e}$ gives the net projected gain or loss around the boundary, while $\gamma_{{\rm g},{\rm e}}$ measures the local deviation from this average.  A sign change of $\gamma_{{\rm g},{\rm e}}$ defines a global dissipative domain wall (GDDW)~\cite{Ma2024}.  We first use the transfer equation to find the boundary envelope, and then set the energy to zero to study its competition with the corner mass domain walls.

\subsection{Transfer generator}\label{subsec:transfer}

To determine how a boundary eigenstate evolves along the directed coordinate $\ell$, we start from the eigenvalue equation $H_{\rm edge}\psi(\ell)=E\psi(\ell)$.  Using Eq.~\eqref{eq:hedge}, it reads
\begin{equation}
\left[-iA_{\rm e}\sz\partial_\ell+m_{\rm e}(\ell)\sy-i\gamma_{\rm e}(\ell)\sz\right]\psi(\ell)=E\psi(\ell).
\end{equation}
This is a first-order differential equation along the boundary.  Multiplying it from the left by $i\sz$ and collecting the derivative on the left gives the transfer equation
\begin{equation}\label{eq:master}
\begin{aligned}
&A_{\rm e}\frac{d\psi}{d\ell}=\mathcal{M}(\ell,E)\psi,\\
&\mathcal{M}(\ell,E)=-\gamma_{\rm e}(\ell)\sI-m_{\rm e}(\ell)\sx+iE\sz.
\end{aligned}
\end{equation}
The matrix $\mathcal M$ is the local transfer generator, and its eigenvalues give the spatial exponents.  On a segment where $m_{\rm e}$ and $\gamma_{\rm e}$ are constant, they are
\begin{equation}\label{eq:lambdas}
\lambda_\pm=-\gamma_{\rm e}\pm\sqrt{m_{\rm e}^{\,2}-E^2}.
\end{equation}
Let $u_\pm$ be the corresponding transfer eigenvectors.  The wave function on this segment can then be written as
\begin{equation}\label{eq:transfer-solution}
\psi(\ell)=\sum_{\nu=\pm}c_\nu e^{\frac{\lambda_\nu(\ell-\ell_0)}{A_{\rm e}}}u_\nu .
\end{equation}
The coefficients $c_\nu$ are fixed by matching neighboring segments.  The real part of $\lambda_\nu/A_{\rm e}$ determines growth or decay as $\ell$ increases.  Their sum is
\begin{equation}\label{eq:trace}
\lambda_++\lambda_-=-2\gamma_{\rm e}.
\end{equation}
Thus, the common part of the two transfer eigenvalues is one half of the trace, $(\lambda_++\lambda_-)/2=-\gamma_{\rm e}$.  On a uniform segment, it gives the common spatial factor $e^{-\gamma_{\rm e}(\ell-\ell_0)/A_{\rm e}}$.  Since $-\gamma_{\rm e}(\ell)\sI$ is scalar in Eq.~\eqref{eq:master}, these local factors multiply along an inhomogeneous boundary profile to give
\begin{equation}\label{eq:skin}
\exp\!\left[-\frac{1}{A_{\rm e}}\int_{\ell_0}^{\ell}\gamma_{\rm e}(\ell')\,d\ell'\right].
\end{equation}
This factor multiplies both transfer components.  The square-root term can still give an additional exponent that depends on the branch.

We first consider a real energy in the propagating part of the edge spectrum, where $|E|>|m_{\rm e}|$.  In this regime, the square root is purely imaginary:
\begin{equation}\label{eq:propagating-root}
\sqrt{m_{\rm e}^{\,2}-E^2}=i\sqrt{E^2-m_{\rm e}^{\,2}} .
\end{equation}
Therefore, $\operatorname{Re}\lambda_+=\operatorname{Re}\lambda_-=-\gamma_{\rm e}$. The mass changes the propagation phase, but not the real transfer exponent. Thus, the two counter-propagating components have the same dissipative envelope.  This happens because the imaginary potential and the kinetic term use the same Pauli matrix; their product is the identity in Eq.~\eqref{eq:master}.

Equivalently, at a fixed real propagating energy on a uniform segment, $\psi\propto e^{ik\ell}$ gives
\begin{equation}
k_\pm=\pm\frac{\sqrt{E^2-m_{\rm e}^{\,2}}}{A_{\rm e}}+i\frac{\gamma_{\rm e}}{A_{\rm e}}.
\end{equation}
The equal imaginary parts of $k_\pm$ give the same decay direction for both components.

The equality of the real exponents requires the propagating conditions above. For complex $E$, the square root in Eq.~\eqref{eq:lambdas} is generally complex and gives opposite real corrections to the two branches.  The same thing happens for a real energy inside the projected edge gap:
\begin{equation}
\begin{aligned}
&\lambda_\pm=-\gamma_{\rm e}\pm\kappa(E),\\
&\kappa(E)=\sqrt{m_{\rm e}^{\,2}-E^2},\qquad |E|<|m_{\rm e}|.
\end{aligned}
\end{equation}
Here, $\kappa(E)$ is real, so the mass adds opposite real exponents.  One branch decays faster and the other decays more slowly, or may even grow.  The common factor in Eq.~\eqref{eq:skin} remains, but it is not the full envelope. This is why the near-zero corner states must be treated separately.

The local transfer equation contains the full projected dissipation.  On a closed boundary, however, its average and its spatial modulation play different roles.  To separate them, we write
\begin{equation}
\psi(\ell)=e^{-\bar\gamma_{\rm e}\ell/A_{\rm e}}\widetilde\psi(\ell).
\end{equation}
Equation~\eqref{eq:master} then becomes
\begin{equation}
A_{\rm e}\partial_\ell\widetilde\psi=\left[-\gamma_{{\rm g},{\rm e}}(\ell)\sI-m_{\rm e}(\ell)\sx+iE\sz\right]\widetilde\psi.
\end{equation}
The average dissipation is removed from the local equation, but not from the closed-boundary problem.  Since the physical wave function satisfies $\psi(\mathcal L)=\psi(0)$, the transformed function satisfies
\begin{equation}\label{eq:twist}
\widetilde\psi(\mathcal L)=e^{+\bar\gamma_{\rm e}\mathcal L/A_{\rm e}}\widetilde\psi(0).
\end{equation}
This non-unitary twist changes the matching condition after one circuit and can therefore change the allowed complex energies.  By itself, however, $\bar\gamma_{\rm e}$ does not select a GDDW or set the density maximum.

The spatial redistribution is controlled by the mean-subtracted part.  Define the corresponding imaginary-gauge factor
\begin{equation}\label{eq:globalskin}
S_{\rm g}(\ell)=e^{-\frac{1}{A_{\rm e}}\int_0^\ell\gamma_{{\rm g},{\rm e}}(\ell')\,d\ell'}.
\end{equation}
Because $\oint\gamma_{{\rm g},{\rm e}}d\ell=0$ by definition, this factor is periodic: $S_{\rm g}(\mathcal L)=S_{\rm g}(0)$.  In the ideal two-component edge theory, it produces the exact similarity relation
\begin{equation}
H_{\rm edge}(\gamma_{{\rm g},{\rm e}})=S_{\rm g}H_{\rm edge}(0)S_{\rm g}^{-1}.
\end{equation}
The transformation is periodic and invertible.  It is scalar in the two-channel space, so it commutes with the position-dependent mass term. When $\bar\gamma_{\rm e}=0$, the full projected dissipation is $\gamma_{{\rm g},{\rm e}}$.  The ideal non-Hermitian edge Hamiltonian is then similar to the Hermitian one: their spectra are the same, but their right eigenstates have different spatial densities.

The direction of the redistribution follows from Eq.~\eqref{eq:globalskin}. The factor $S_{\rm g}$ grows where $\gamma_{{\rm g},{\rm e}}<0$ and decays where $\gamma_{{\rm g},{\rm e}}>0$.  Its maximum is therefore at the sign change from negative to positive. This sign change is the accumulating GDDW.  The opposite sign change is the depleting GDDW.  For a balanced profile, $\gamma_{{\rm g},{\rm e}}=\gamma_{\rm e}$. 

Now, we can obtain the localization length of the common propagating envelope.  Let $d\geq0$ denote the local boundary distance measured away from an accumulating GDDW,  then density decreases as $e^{-2|\gamma_{{\rm g},{\rm e}}|d/A_{\rm e}}$, there is
\begin{equation}\label{eq:lengthdefinitions}
\begin{aligned}
&\rho(d)=|\widetilde\psi(d)|^2\propto e^{-d/\xi},\\
&\xi\simeq\frac{A_{\rm e}}{2|\gamma_{{\rm g},{\rm e}}|}.
\end{aligned}
\end{equation}
Equation~\eqref{eq:lengthdefinitions} describes only the relative common envelope of propagating states.  It is not the full localization length of a subgap corner state, for which the mass contribution must also be included.

A nonzero mean does not simply add the same imaginary constant to every eigenvalue, because $E$ enters the transfer generator as the matrix term $iE\sz$.  The full two-channel matching problem must be solved.  A finite lattice realizes the ideal projected theory only approximately.  Corner mixing, residual Pauli components, and coupling to higher-energy states can produce additional complex shifts or splittings.  These finite-size deviations are assessed below using a fixed-energy Feshbach reduction, which retains the discrete boundary geometry, corner coupling, perimeter closure, and the interior self-energy omitted from the ideal edge theory.

\subsection{Zero-energy envelopes}\label{subsec:zero}

The preceding propagating-state result isolates the common dissipative envelope.  A corner state lies inside the projected edge gap and already has a mass-induced Jackiw--Rebbi envelope in the Hermitian limit.  Its mass contribution must therefore be retained.  Setting $E=0$ in Eq.~\eqref{eq:master} gives
\begin{equation}
A_{\rm e}\partial_\ell\psi=\left[-\gamma_{\rm e}(\ell)\sI-m_{\rm e}(\ell)\sx\right]\psi.
\end{equation}
The two matrices in this equation are diagonal in the $\sx$ basis.  We therefore write $\psi(\ell)=g_s(\ell)\ket{s}_{\sx}$, where $s=\pm1$ and $\sx\ket{s}_{\sx}=s\ket{s}_{\sx}$.  Each component of the original physical wave function obeys
\begin{equation}\label{eq:Vsraw}
A_{\rm e}\partial_\ell g_s=\left[-\gamma_{\rm e}(\ell)-s\,m_{\rm e}(\ell)\right]g_s.
\end{equation}
Using the same mean subtraction as above, $g_s=e^{-\bar\gamma_{\rm e}\ell/A_{\rm e}}\widetilde g_s$, the relative zero-energy envelope satisfies
\begin{equation}\label{eq:Vs}
A_{\rm e}\partial_\ell \widetilde g_s=V_s(\ell)\widetilde g_s,\qquad V_s(\ell)=-\gamma_{{\rm g},{\rm e}}(\ell)-s\,m_{\rm e}(\ell),
\end{equation}
with the solution
\begin{equation}\label{eq:gsolution}
\widetilde g_s(\ell)=\widetilde g_s(\ell_0)e^{\frac{1}{A_{\rm e}}\int_{\ell_0}^{\ell}V_s(\ell')\,d\ell'}.
\end{equation}
For $\bar\gamma_{\rm e}\neq0$, $\widetilde g_s$ obeys the twisted boundary condition in Eq.~\eqref{eq:twist}.

The function $V_s$ combines the two sources of localization.  The relative envelope grows along increasing $\ell$ when $V_s>0$, and it decays when $V_s<0$.  A local maximum occurs where $V_s$ changes from positive to negative.  In the Hermitian limit, $V_s=-s m_{\rm e}$, so these changes are the familiar MDWs that bind Jackiw--Rebbi corner states.  The global dissipation adds the same term $-\gamma_{{\rm g},{\rm e}}$ to both $\sx$ channels.  More explicitly, the logarithmic amplitude growth rate is
\begin{equation}
\partial_\ell\ln|\widetilde g_s|=\frac{-\gamma_{{\rm g},{\rm e}}-s m_{\rm e}}{A_{\rm e}}.
\end{equation}
This equation makes the competition transparent.  When the mass term is larger, the density remains concentrated near the geometric corners.  As the global dissipation increases, it shifts the balance of growth and decay toward the accumulating GDDWs.  For piecewise-constant profiles, the leading local crossover occurs when
\begin{equation}
|\gamma_{{\rm g},{\rm e}}|\sim |m_{\rm e}|.
\end{equation}

The local growth rule is not the only condition on a closed boundary.  Within the ideal periodic two-channel edge model, an exact zero-energy solution must return to its original value after one circuit. Integrating Eq.~\eqref{eq:Vsraw} around the boundary gives the condition
\begin{equation}\label{eq:periodiczero}
\oint\left[-\gamma_{\rm e}(\ell)-s\,m_{\rm e}(\ell)\right]d\ell=-\bar\gamma_{\rm e}\mathcal L-s\oint m_{\rm e}(\ell)d\ell=0.
\end{equation}
The mean-subtracted part integrates to zero by definition.  For the symmetric square mass profile, $\oint m_{\rm e}d\ell=0$ as well.  A strictly periodic zero mode in one $\sx$ channel therefore requires $\bar\gamma_{\rm e}=0$. Otherwise it can move away from zero energy and mix with the other channel. On a finite lattice, separated domain-wall states can also hybridize.  We therefore refer to them as a low-energy domain-wall-state subspace, rather than as exact zero modes at every sign change.

\subsection{Symmetry and dissipation channels}\label{subsec:channels}

The kinetic-aligned edge Hamiltonian retains chiral symmetry,
\begin{equation}\label{eq:chiral}
\sx H_{\rm edge}\sx=-H_{\rm edge},
\end{equation}
because $\sx$ anticommutes with both $\sz$ and $\sy$.  This relation holds for any real profile $\gamma_{\rm e}(\ell)$.  It pairs $E$ with $-E$, but it does not make the spectrum real.  Reality follows from the stronger similarity relation above.  It requires both $\bar\gamma_{\rm e}=0$ and alignment with the kinetic Pauli matrix; a zero average is not enough for a general boundary operator.

We next ask a broader question: is momentum complexification by itself enough to localize the two edge components?  To compare different possibilities, we replace $-i\gamma_{\rm e}\sz$ by $-i\gamma_{\rm e}\tauh{\alpha}$, where $\alpha=0,x,y,z$.  For
\begin{equation}
H=-iA_{\rm e}\partial_\ell\sz+m_{\rm e}\sy-i\gamma_{\rm e}\tauh{\alpha},
\end{equation}
the dissipative part of the transfer generator becomes
\begin{equation}\label{eq:channels}
\mathcal{M}_\alpha=-\gamma_{\rm e}(\sz\tauh{\alpha})-m_{\rm e}\sx+iE\sz.
\end{equation}
The result depends on the matrix product $\sz\tauh{\alpha}$, not only on whether the wave number becomes complex.

For the kinetic-aligned channel, $\tauh{\alpha}=\sz$, Eq.~\eqref{eq:channels} reduces to Eq.~\eqref{eq:master} and reproduces the common envelope and same-side localization derived above.

We now use scalar boundary dissipation as a control for the kinetic-aligned result.  For the Hamiltonian channel $-i\gamma_{\rm e}\sI$, the transfer generator contains $-\gamma_{\rm e}\sz$.  At $m_{\rm e}=0$ and a fixed real energy, the two decoupled $\sz$ components have the local wave numbers
\begin{equation}
k_s=\frac{sE}{A_{\rm e}}+i\frac{s\gamma_{\rm e}}{A_{\rm e}},\qquad s=\pm1.
\end{equation}
In the massless limit, the two kinetic components are independent, and scalar dissipation would localize them at opposite GDDWs.  The mass term $-m_{\rm e}\sx$ does not commute with $-\gamma_{\rm e}\sz$.  Finite $m_{\rm e}$ mixes the two components, so scalar dissipation no longer supports two independent chiral envelopes and does not generically produce the common GDDW accumulation.  Complex momentum alone therefore does not fix whether or where localization occurs.  The outcome also depends on the matrix structure of the dissipation and its relation to the edge mass.

The remaining Pauli channels provide further contrasts.  The $\sx$ channel gives $-i\gamma_{\rm e}\sy$ in the transfer generator, while the $\sy$ channel gives $+i\gamma_{\rm e}\sx$.  Both terms are traceless.  They modify channel mixing or the effective mass instead of adding a common scalar exponent.  They may therefore reshape the localization in a state-dependent way, but they do not generate the branch-independent envelope of the kinetic-aligned channel.  In the present lattice model, the alignment of the spin-dependent dissipation with the kinetic Pauli matrix follows from spin--velocity locking.

%=================================================================
\section{NUMERICAL RESULTS}
\label{sec:numerics}
%=================================================================
We test the edge theory by exact diagonalization of a finite square lattice.  Using $\ell$ and $\mathcal L$ as defined in Sec.~\ref{sec:edge}, with $\ell=0$ at the lower-left corner, we choose the bare boundary dissipation profile
\begin{equation}\label{eq:separatedprofile}
\begin{aligned}
&\gamma(\ell)=\begin{cases}
+\gamma,&0\leq\ell<\mathcal L/8\ \text{or}\ 5\mathcal L/8\leq\ell<\mathcal L,\\
-\gamma,&\mathcal L/8\leq\ell<5\mathcal L/8,
\end{cases}
\end{aligned}
\end{equation} 
where $\gamma>0$, and the profile is shown schematically in the inset of Fig.~\ref{fig:analyticnumeric}(a).

\begin{figure}[!t]
\centering
\includegraphics[width=0.98\columnwidth]{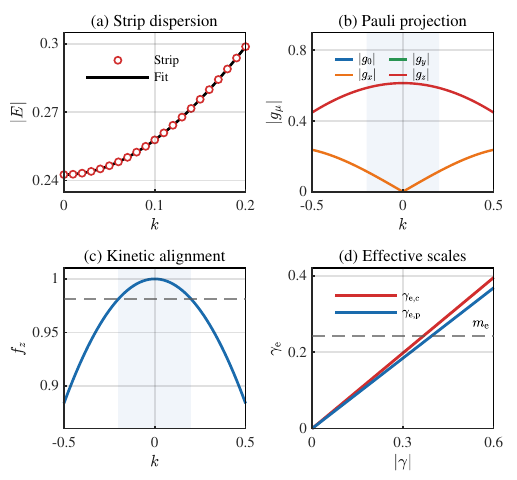}
\caption{Effective parameters fixed from Hermitian calculations for $A=B=1$, $M=1$, and $\Lambda=0.5$.  (a) Low-energy strip spectrum and the fit $|E|^2=m_{\rm e}^2+A_{\rm e}^2k^2$.  (b) Magnitudes $|g_\mu(k)|$ in the Pauli decomposition of the outer-row $s_z$ operator in the kinetic gauge.  (c) Kinetic-aligned fraction $f_z$; the dashed line marks $f_z=0.9812$.  The shaded range in (b) and (c) is $|k|\leq0.2$.  (d) Corner-state outer-layer exposure scale, $\gamma_{{\rm e,c}}^{(0)}=w_0|\gamma|$ (red), and projected effective dissipation of the propagating doublet, $\gamma_{{\rm e,p}}^{(0)}=|g_z(0)||\gamma|$ (blue); the dashed line is $m_{\rm e}$.  All fit parameters and calibration factors are obtained at $\gamma=0$.}
\label{fig:effectiveparameters}
\end{figure}

\begin{figure*}[!t]
\centering
\includegraphics[width=0.98\textwidth,trim=0 14bp 9bp 18bp,clip]{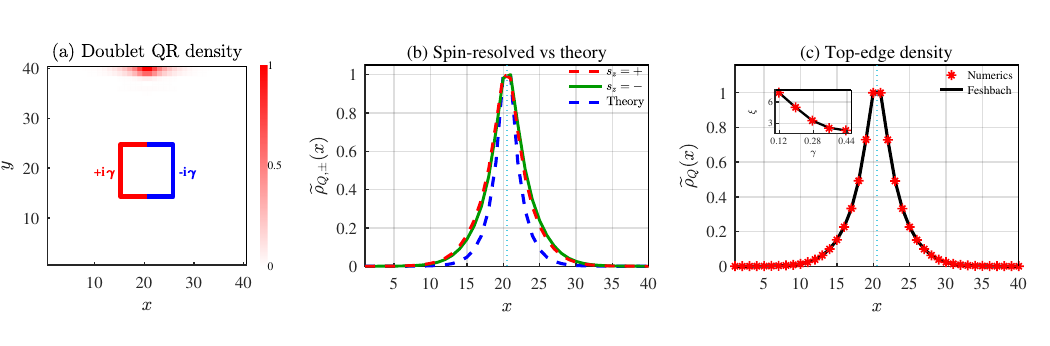}
\caption{Propagating boundary states for $L=40$ at $\gamma_{\rm e}=0.2668$.  (a) Numerics QR-projector density of the tracked positive-energy double, and the inset shows the dissipation profile.  (b) The $s_z=+$ (red dashed) and $s_z=-$ (green solid) parts of the upper-edge QR-projector density, together with the envelope $|S_{\rm g}(\ell)|^2$ from Eq.~\eqref{eq:globalskin} (blue dashed).  All three profiles peak at the GDDW marked by the cyan dotted line.  (c) The numucial (red stars) and the fixed-energy Feshbach reduction (black solid) at $y=L$.  The inset compares the localization lengths as $\gamma$ is varied, with $\gamma_{\rm e}=|g_z(0)|\gamma$.  Here $\widetilde{\rho}_Q(x)$ is the peak-normalized QR-projector density, and $\widetilde{\rho}_{Q,\pm}(x)$ are its spin-resolved parts.  The lattice parameters are the same as in Fig.~\ref{fig:effectiveparameters}.}
\label{fig:analyticnumeric}
\end{figure*}
\begin{figure}[!t]
\includegraphics[width=0.42\textwidth,trim=0 7bp 8bp 3bp,clip]{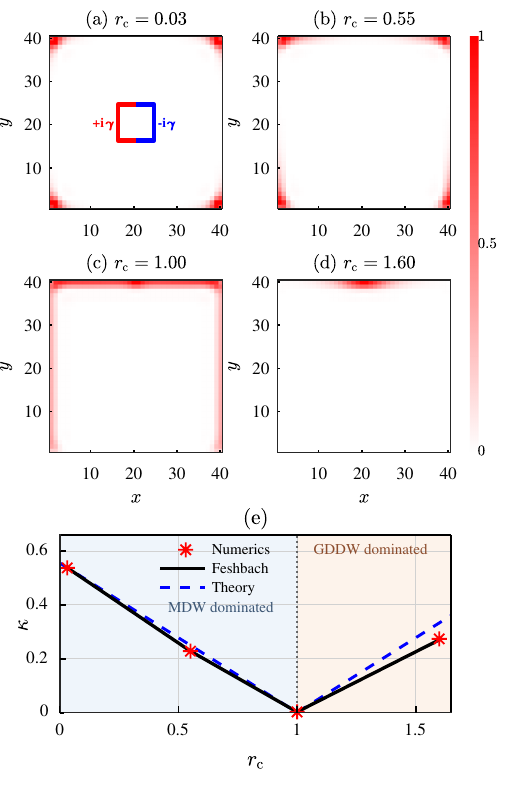}
\caption{Redistribution of the QR-projector density of the tracked corner-state subspace for $L=40$.  (a)--(d) QR-projector densities at $r_{\rm c}=0.03,0.55,1.00,1.60$.  The inset shows the dissipation profile.  (e) Upper-half-edge inverse density localization length $\kappa$ versus $r_{\rm c}=w_0\gamma/m_{\rm e}$:  numerics (red stars), fixed-energy Feshbach reduction (black solid), and the edge theory  from Eqs.~\eqref{eq:Vs} and \eqref{eq:gsolution} (blue dashed).  The dotted line at $r_{\rm c}=1$ and the shaded regions mark the leading crossover from MDW-dominated to GDDW-dominated localization.  The lattice parameters are the same as in Fig.~\ref{fig:effectiveparameters}.}
\label{fig:migration3d}
\end{figure}

Figure~\ref{fig:effectiveparameters} summarizes the extracted effective boundary parameters.  Fitting the low-energy strip dispersion to
\begin{equation}\label{eq:stripfit}
|E(k)|=\sqrt{m_{\rm e}^{\,2}+A_{\rm e}^{\,2}k^2},
\end{equation}
gives $m_{\rm e}=0.242536$ and $A_{\rm e}=0.874$, as shown in Fig.~\ref{fig:effectiveparameters}(a). For the finite lattice, we write
\begin{equation}\label{eq:Gdef}
\begin{aligned}
&H(\gamma)=H(0)+i\gamma G,\\
&G=\sum_{\bm r\in\partial\Omega}\zeta(\ell_{\bm r})|\bm r\rangle\langle\bm r|\otimes Q_{0z},\\
&Q_{0z}\equiv\sigma_0s_z.
\end{aligned}
\end{equation}
Here, $\ell_{\bm r}$ is the coordinate of boundary site $\bm r$, and $\zeta(\ell)=\gamma(\ell)/\gamma=\pm1$ is fixed by Eq.~\eqref{eq:separatedprofile}, and $G$ acts only on the outermost-layer sites.  The corner-state subspace and the propagating edge doublet are calibrated separately.

For the propagating doublet, effective dissipation is obtained by projecting the outermost-row spin operator onto the two-state edge subspace; Appendix~\ref{app:projection} gives the basis and projection procedure.  The resulting Pauli components are shown in Fig.~\ref{fig:effectiveparameters}(b).  Near $k=0$, $g_z$ is dominant, $g_0=g_y=0$, and $g_x$ is the leading finite-$k$ correction. Over the low-energy momentum range $|k|\leq0.2$, the kinetic-aligned fraction $f_z=|g_z|/(\sum_\mu|g_\mu|^2)^{1/2}$ satisfies $f_z\geq0.9812$, as shown in Fig.~\ref{fig:effectiveparameters}(c).  This shows that the projected dissipation acts mainly through the kinetic channel. At $k=0$, this projection gives $\gamma_{{\rm e,p}}^{(0)}=|g_z(0)|\gamma$ for the propagating doublet.  All four strip calculations give the same coefficient, $|g_z(0)|=0.614288$, so the same projected scale applies to all four edges.  For the corner-state subspace, $w_0=0.659223$ is the average weight on the outermost layer at the Hermitian point.  We use $\gamma_{{\rm e,c}}^{(0)}=w_0\gamma$ only as an outer-layer exposure scale.  Figure~\ref{fig:effectiveparameters}(d) compares these two scales with the edge mass $m_{\rm e}$.

\begin{figure}[!t]
\centering
\includegraphics[width=\columnwidth]{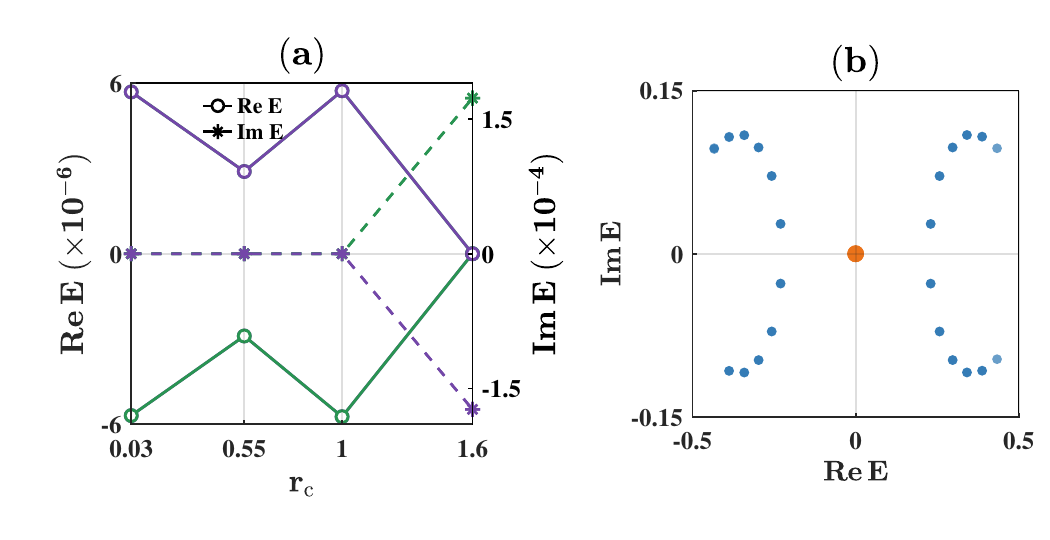}
\caption{Spectrum of the four tracked corner-derived states for $L=40$.  (a) Real and imaginary parts of their eigenvalues at $r_{\rm c}=0.03,0.55,1.00,1.60$.  (b) The 48 eigenvalues nearest zero at $r_{\rm c}=1.60$, with the four tracked eigenvalues highlighted.  The lattice parameters are the same as in Fig.~\ref{fig:effectiveparameters}.}
\label{fig:complex}
\vspace{0.5\baselineskip}
\includegraphics[width=0.98\columnwidth]{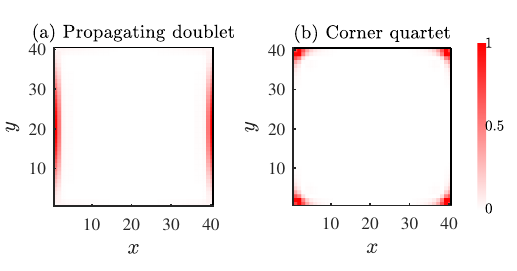}
\caption{Scalar boundary dissipation for $L=40$ at $\gamma=0.5887$.  Peak-normalized QR-projector densities are shown for (a) the tracked positive-energy propagating doublet and (b) the tracked corner-state quartet.  The scalar operator $Q_{00}\equiv\sigma_0s_0$ follows the balanced separated profile in both panels.  Each panel is normalized by its own maximum.  The lattice parameters are the same as in Fig.~\ref{fig:effectiveparameters}.}
\label{fig:channel}
\end{figure}

We first test the propagating-state prediction on the finite lattice.  Nearly degenerate right eigenvectors can mix in a non-Hermitian calculation, so we use the QR-projector density of a tracked subspace rather than the density of a single right eigenvector~\cite{GolubVanLoan2013,StewartSun1990}.  Appendix~\ref{app:subspace} gives the definition and tracking procedure.  Starting from the positive-energy Hermitian propagating doublet $Q_{{\rm P},+}^{(0)}$, we use the balanced profile in Eq.~\eqref{eq:separatedprofile}, for which $\bar\gamma_{\rm e}=0$.  The accumulating and depleting GDDWs lie at the midpoints of the upper and lower edges, respectively.  Equations~\eqref{eq:globalskin} and \eqref{eq:lengthdefinitions} then predict that both kinetic channels share a density envelope centered at the upper GDDW, with a localization length that decreases as the dissipation increases.

Figure~\ref{fig:analyticnumeric}(a) confirms that the numerical QR-projector density peaks at the upper-edge GDDW.  In Fig.~\ref{fig:analyticnumeric}(b), the $s_z=\pm$ parts are obtained from the same QR projector using $P_{s,\pm}=(s_0\pm s_z)/2$.  Since $P_{\rm edge}s_zP_{\rm edge}=-\sz$, they represent the two counter-propagating kinetic channels.  Both parts peak at the same GDDW and follow the decay trend of $|S_{\rm g}(\ell)|^2$, although the edge-theory envelope is slightly narrower because of finite-lattice spin mixing and projection corrections.

To include these finite-lattice corrections, we use a Feshbach reduction at a fixed Hermitian reference energy~\cite{Feshbach1958}.  It retains the discrete boundary and its coupling to the interior; Appendix~\ref{app:feshbach} gives the construction.  Figure~\ref{fig:analyticnumeric}(c) and its inset show that the reduced density and localization length agree closely with the full-lattice results. Therefore,
the fixed-energy reduction captures the full-lattice result well.

Next, turning to the corner states. For corner states, the dissipative envelope competes with the Jackiw–Rebbi confinement. We characterize this competition by
\begin{equation*}
r_{\rm c}=\frac{\gamma_{{\rm e,c}}^{(0)}}{m_{\rm e}}=\frac{w_0\gamma}{m_{\rm e}},
\end{equation*}
and examine how the corner-state subspace is redistributed as $r_{\rm c}$ increases. For a small $r_{\rm c}$, dissipation is weak, so the density remains localized at the four mass domain walls. Figures~\ref{fig:migration3d}(a) and \ref{fig:migration3d}(b) show this MDW-dominated regime.  As the dissipation increases, the density leaves the MDWs and moves toward the accumulating GDDW.  At $r_{\rm c}=1$, the dissipation and mass scales are comparable, so the density spreads along the boundary [Fig.~\ref{fig:migration3d}(c)].  For $r_{\rm c}>1$, the GDDW becomes dominant, and the density is localized at the accumulating GDDW, as shown at $r_{\rm c}=1.60$ in Fig.~\ref{fig:migration3d}(d).  Equation~\eqref{eq:Vs} explains this behavior.  Figure~\ref{fig:migration3d}(e) compares the inverse density localization length $\kappa$ from the full-lattice numerics, the effective edge theory, and the fixed-energy Feshbach reduction.  The three results agree well and show the competition between MDW and GDDW localization.

Figure~\ref{fig:complex} provides an independent spectral check of the same tracked subspace.  In Fig.~\ref{fig:complex}(a), the real parts remain close to zero, while the imaginary parts split at strong dissipation, indicating different growth or decay rates.  At $r_{\rm c}=1.60$, the four tracked states remain near zero and are separated from the next state by $0.231$ [Fig.~\ref{fig:complex}(b)].  The subspace can therefore still be identified.

Finally, we test the channel dependence predicted in Sec.~\ref{subsec:channels}.  We apply the same separated profile through the scalar boundary operator $Q_{00}\equiv\sigma_0s_0$ and track the positive-energy propagating doublet from $Q_{{\rm P},+}^{(0)}$ and the corner-state quartet from $Q_{\rm C}^{(0)}$.  At this strong scalar dissipation, the propagating-doublet density peaks at the midpoints of the left and right edges, away from both GDDWs [Fig.~\ref{fig:channel}(a)], while the corner-state density remains at the four MDWs [Fig.~\ref{fig:channel}(b)].  Although scalar dissipation can give complex boundary wave numbers, the finite mass mixes the two kinetic components and prevents a common GDDW envelope.  The absence of a propagating-density peak at either GDDW also rules out two branch-resolved accumulations at opposite GDDWs.  Thus, the matrix structure of the boundary dissipation is essential.

\section{Conclusion}\label{sec:conclusion}

Inhomogeneous, spin-dependent boundary dissipation provides a way to control boundary-state localization in a BHZ-type SOTI. The effective edge theory explains the underlying mechanism. In conventional chiral-skin theory, edge modes of opposite chirality localize at opposite GDDWs. Here, the projected dissipation reverses sign together with the propagation velocity, giving the two counter-propagating components the same real transfer exponent and localizing both at the same GDDW. These components are also coupled by an edge mass, which forms MDWs and binds Jackiw--Rebbi corner states. The competition between mass confinement and the common dissipative envelope shifts the corner-state weight from the geometric corners toward the accumulating GDDW as the dissipation increases.

Numerical results qualitatively confirm the effective edge theory. The fixed-energy Feshbach reduction agrees closely with the numerical tracked QR density and localization lengths, and accounts for the finite-size deviations from the ideal edge theory. Because the dissipation acts only on the boundary, the periodic bulk Hamiltonian and its topology remain unchanged.  Therefore, inhomogeneous boundary dissipation thus controls boundary-state localization without changing the bulk topology.

Topological corner modes and non-Hermitian higher-order skin effects have been realized in topolectrical circuits~\cite{Imhof2018,Zou2021}. These results suggest that an RLC circuit could be used to test our mechanism. An entirely passive implementation is also possible. A uniform loss can be added to all circuit nodes, while boundary resistors produce different loss rates for the two effective spin components. If the uniform loss is larger than the boundary modulation, both loss rates remain positive. The uniform part only shifts the complex spectrum and does not change the boundary-state density.

\begin{acknowledgments}
We are grateful to Kui Cao and Zheng Wei for their valuable suggestions regarding the manuscript.This work was supported by the National Natural Science Foundation of China (Grants No.~11974053 and No.~12174030) and the National Key R\&D Program of China (Grant No.~2023YFA1406704).
\end{acknowledgments}

\appendix

\section{Boundary projection and sign conventions}\label{app:projection}

This appendix fixes the boundary coordinate, the order of the two edge states, and the projected parameters.

\subsection{Bottom-edge convention}

We use the bottom edge to fix the convention, considering a system occupying the region $y>0$, on the bottom edge, we take $k_\ell=k_x$.  Near the $\Gamma$ point, Eq.~\eqref{eq:bulk} becomes~\cite{Yan2019}
\begin{equation}\label{eq:bhzcontinuum}
\begin{aligned}
H\simeq{}&[M-B(k_\ell^2+k_y^2)]\sigma_z+A k_\ell\sigma_xs_z+A k_y\sigma_y\\
&-\frac{\Lambda}{2}(k_\ell^2-k_y^2)\sigma_xs_x.
\end{aligned}
\end{equation}
Here, the unimportant higher-order term $k_\ell^2$ is neglected.  At $\Lambda=0$, there are two states on the bottom edge, they have the same wave-function  $f_{\rm B}(y)$ perpendicular to the boundary.
\begin{equation}\label{eq:bottom-envelope}
\begin{aligned}
&f_{\rm B}(y)=\mathcal N\left(e^{-\lambda_1 y}-e^{-\lambda_2 y}\right),\\
&\lambda_{1,2}=\frac{A\pm\sqrt{A^2-4BM}}{2B}.
\end{aligned}
\end{equation}
where $\mathcal N$ is a normalization constant, the orbital part is
\begin{equation}
|\sigma_x=-1\rangle=\frac{|\sigma_z=+1\rangle-|\sigma_z=-1\rangle}{\sqrt2},
\end{equation}
here, the orbital part is fixed to $\sigma_x=-1$, while the spin can still be either up or down. We use the following order for the microscopic basis
\begin{equation}
\mathcal B_{\rm mic}=\bigl(|+,+\rangle,|+,-\rangle,|-,+\rangle,|-,-\rangle\bigr),
\end{equation}
On the bottom edge, we choose
\begin{equation}\label{eq:bottominternalbasis}
\begin{aligned}
&\mathcal U_{\rm B}=\left(|\sigma_x=-1,\downarrow\rangle,\,i|\sigma_x=-1,\uparrow\rangle\right)\\
&=\frac{1}{\sqrt2}\begin{pmatrix}
0&i\\
1&0\\
0&-i\\
-1&0
\end{pmatrix},\qquad U_{\rm B}(y)=f_{\rm B}(y)\mathcal U_{\rm B}.
\end{aligned}
\end{equation}
The first state moves along $+\ell$ with velocity $+A$. The second state moves in the opposite direction with velocity $-A$. The $i$ does not change the physical result, and it only lets us write the projected mass term as $\hat{\tau}_y$. In this basis, these operators become
\begin{equation}\label{eq:bottombasischecks}
\begin{aligned}
&\mathcal U_{\rm B}^\dagger\mathcal U_{\rm B}=I_2,\qquad \mathcal U_{\rm B}^\dagger s_z\mathcal U_{\rm B}=-\sz,\\
&\mathcal U_{\rm B}^\dagger(\sigma_xs_z)\mathcal U_{\rm B}=\sz,\qquad \mathcal U_{\rm B}^\dagger(\sigma_xs_x)\mathcal U_{\rm B}=\sy.
\end{aligned}
\end{equation}
The bottom-edge Hamiltonian therefore takes the form
\begin{equation}\label{eq:bottomedgehamiltonian}
H_{{\rm B},{\rm edge}}(k_\ell)=A k_\ell\sz+m_0\sy-i\gamma_{\rm e}(\ell)\sz,\qquad m_0=\frac{\Lambda M}{2B}.
\end{equation}
The last term contains the effective dissipation $\gamma_{\rm e}$. We obtain it by projecting the outermost-layer spin operator $P_{\rm out}s_z$ onto the edge states.
\subsection{Using the same convention on all four edges}

The first part defines the convention on the bottom edge.  Now, applying the same convention to four edges of the lattice.  In the lattice model, an edge state does not always have a fixed $s_z$,  so, we use numerical calculations to find the two basis states on each edge. First, finding the two states on this edge, then,  putting the positive-velocity state first and the negative-velocity state second. At a fixed $k_\ell$, a Hermitian strip has two physical edges, and each edge has two low-energy states, so we take four states in total, then these four orthonormal states can be written 
\begin{equation}
\Phi_j=(|\phi_{j,1}\rangle,|\phi_{j,2}\rangle,|\phi_{j,3}\rangle,|\phi_{j,4}\rangle)
\end{equation}
To select the two states on the chosen edge, $P_{\rm targ}$ is taken as the projector onto the outermost row of that edge, and its matrix in the four-state space is
\begin{equation}
M^{\rm pos}_j=\Phi_j^\dagger P_{\rm targ}\Phi_j.
\end{equation}
The two eigenvectors of $M^{\rm pos}_j$ with the largest eigenvalues are kept because they give the two linear combinations with the largest weight on the chosen edge, and after these normalized eigenvectors are put into $W^{\rm pos}_j$, the corresponding edge states are
\begin{equation}
\widetilde U_j=\Phi_jW^{\rm pos}_j,\qquad (W^{\rm pos}_j)^\dagger W^{\rm pos}_j=I_2.
\end{equation}
The matrix $\widetilde U_j$ therefore contains the two low-energy basis states associated with the chosen edge, while their velocity order is not fixed yet.

To fix this order, the velocity operator is first projected onto the two states, and the trace of the resulting matrix is removed:
\begin{equation}\label{eq:projectedvelocity}
V_j=\widetilde U_j^\dagger\frac{\partial H_{{\rm strip},j}}{\partial k_\ell}\widetilde U_j,\qquad \widetilde V_j=V_j-\frac{\tr V_j}{2}I_2.
\end{equation}
The trace gives the same velocity shift to both states and does not change the eigenvectors, so the remaining matrix is diagonalized:
\begin{equation}
\widetilde V_j|w_{j,\pm}\rangle=\pm v_j|w_{j,\pm}\rangle
\end{equation}
The positive-velocity state is placed first and the negative-velocity state second, giving the momentum-dependent basis
\begin{equation}\label{eq:kineticbasis}
U_j(k_\ell)=\widetilde U_j(k_\ell)(|w_{j,+}\rangle,|w_{j,-}\rangle).
\end{equation}
Only the two states on the same edge are mixed when forming $U_j$, but their phases are still free; these phases are chosen so that the projected mass is written with the same $\sy$ matrix on every edge, and a low-energy state on edge $j$ is then expressed as
\begin{equation}\label{eq:localedgeexpansion}
|\Psi_j(\ell,n_\perp)\rangle\simeq U_j(n_\perp)|\chi_j(\ell)\rangle,\qquad |\chi_j\rangle=(a_{j,+},a_{j,-})^{\mathsf T}.
\end{equation}
A single coordinate $\ell$ is taken counterclockwise around the square, and on every edge the labels $+$ and $-$ mean positive and negative velocity, respectively; the momenta are then
\begin{equation}\label{eq:fourkell}
\begin{aligned}
&k_\ell=\begin{cases}
+k_x,&\text{bottom},\\
+k_y,&\text{right},\\
-k_x,&\text{top},\\
-k_y,&\text{left}.
\end{cases}
\end{aligned}
\end{equation}
With this choice, $\sz=+1$ always means motion in the counterclockwise direction, so the kinetic term has the form $-iA_{\rm e}\partial_\ell\sz$ on every edge.  If $k_\ell$ is reversed or the two columns of $U_j$ are exchanged, the sign of $\sz$ is also reversed, so the direction of $\ell$ and the order of the two states must be fixed before the sign of the projected coefficient is determined.

In the ideal continuum model with $\Lambda=0$, the same relation between spin and velocity is found on all four edges:
\begin{equation}\label{eq:signedprojection}
U_j^\dagger s_zU_j=-\sz.
\end{equation}
In the lattice model, however, the boundary dissipation acts only on the outermost sites, so  its projection onto the two edge states is
\begin{equation}\label{eq:fourprojectedloss}
\begin{aligned}
&\Gamma_j(k_\ell)=U_j^\dagger(k_\ell)P_{\rm out}s_zU_j(k_\ell)=\sum_{\mu=0,x,y,z}g_{\mu,j}(k_\ell)\tau_\mu,\\
&g_{\mu,j}(k_\ell)=\frac12\tr\left[\Gamma_j(k_\ell)\tau_\mu\right].
\end{aligned}
\end{equation}
where $P_{\rm out}$ is boundary projection operator.  At low energy, the $g_{z,j}$ term is much larger than the other terms, so the projected dissipation becomes
\begin{equation}
i\gamma(\ell)\Gamma_j\simeq-i\gamma_{\rm e}(\ell)\sz,\qquad \gamma_{\rm e}(\ell)=-\gamma(\ell)g_{z,j}.
\end{equation}
With the velocity order in Eq.~\eqref{eq:kineticbasis}, the four strip calculations give
\begin{equation}\label{eq:fouredgenumericprojection}
\Gamma_j(0)=-0.614288\,\sz,\qquad j={\rm B,R,T,L},
\end{equation}
Thus, $g_{z,j}(0)=-0.614288<0$, so $\gamma_{\rm e}(\ell)$ has the same sign as the bare profile $\gamma(\ell)$, while its magnitude is reduced by the projection.
Their magnitude can be written as $0.614288=0.633193\times0.970143$, where $0.633193$ is the probability on the outermost row and $0.970143$ is the spin polarization on that row.  The same sign is obtained because the same counterclockwise coordinate and velocity order are used for every strip.

The same basis also gives $P_js_zP_j=-\sz$ on the other three edges, and the projected mass is written with the same $\sy$ matrix on every edge.  In the order bottom, right, top, and left, the masses are $(+m_0,-m_0,+m_0,-m_0)$, so each change of sign forms a mass domain wall.

The continuum result is not exact for a finite lattice because finite $\Lambda$, boundary details, and corner mixing give corrections.  The finite-lattice calculations therefore use the effective parameters extracted in Fig.~\ref{fig:effectiveparameters}.

At a corner, the bases on the two neighboring edges are related by a unitary transformation, and a different local gauge may reverse both $\sz$ and its coefficient.  Their product, and therefore the projected operator, remains unchanged, so the gauge choice cannot change the GDDW where the boundary states accumulate.

\section{Subspace tracking}\label{app:subspace}

The $d$ selected right eigenvectors are collected in
\begin{equation}\label{eq:psirappendix}
\Psi_R=\left(|\psi_1^R\rangle,\ldots,|\psi_d^R\rangle\right)\in\mathbb C^{D\times d},\qquad D=4L^2.
\end{equation}
For linearly independent columns, thin QR gives~\cite{GolubVanLoan2013}
\begin{equation}\label{eq:thinQRappendix}
\Psi_R=Q_R\mathcal R,\qquad Q_R^\dagger Q_R=I_d,
\end{equation}
Thus, $Q_R$ is an orthonormal basis for the same selected right subspace.  Its basis-independent projector is
\begin{equation}\label{eq:projectorappendix}
\Pi_R=Q_RQ_R^\dagger.
\end{equation}
It is unchanged by normalization, ordering, or nonsingular mixing of the selected eigenvectors.  We therefore define the density from $\Pi_R$ rather than from a single right eigenvector.

The spatial density is obtained from the diagonal of this projector, with the four internal states summed at each site:
\begin{equation}\label{eq:qrdensityappendix}
\rho_Q(\bm r)=\sum_{\alpha=1}^{4}\langle\bm r,\alpha|\Pi_R|\bm r,\alpha\rangle.
\end{equation}
Next, we define the two starting subspaces at $\gamma=0$.

For the corner-state subspace, the full open-boundary lattice is diagonalized and the low-energy eigenpairs are sorted by $|E|$, so the four modes closest to zero define
\begin{equation}\label{eq:cornerreference}
Q_{\rm C}^{(0)}=\operatorname{qr}\left(|\psi_1^{(0)}\rangle,\ldots,|\psi_4^{(0)}\rangle\right).
\end{equation}
Their QR density lies mainly near the boundary and the corners, so we use $Q_{\rm C}^{(0)}$ as the starting corner-state subspace.

For the propagating subspace, the two closest positive-energy modes with large boundary weight are selected, and their raw matrix is denoted by $\Psi_{{\rm P},+}^{(0)}$.  Because a small numerical component may still lie in the corner-state subspace, it is removed before thin QR is applied:
\begin{equation}\label{eq:propreference}
\begin{aligned}
&\widetilde\Psi_{{\rm P},+}^{(0)}=(I-\Pi_{\rm C}^{(0)})\Psi_{{\rm P},+}^{(0)},\\
&Q_{{\rm P},+}^{(0)}=\operatorname{qr}\left(\widetilde\Psi_{{\rm P},+}^{(0)}\right).
\end{aligned}
\end{equation}
After the two starting subspaces are fixed, they are tracked as the dissipation is increased.  At each $\gamma$, each low-energy subspace is compared with the previously selected subspace $Q_n$, and its overlap is calculated from the QR basis $Q_{\mathcal S}$~\cite{StewartSun1990}:
\begin{equation}\label{eq:overlaptraceappendix}
\mathcal O_n(\mathcal S)=\frac1d\left\|Q_n^\dagger Q_{\mathcal S}\right\|_F^2,\qquad 0\leq\mathcal O_n\leq1.
\end{equation}
The group with the largest overlap is used as $Q_{n+1}$.

\section{Fixed-energy Feshbach reduction}\label{app:feshbach}

We use a fixed-energy Feshbach reduction for the propagating and corner-state comparisons~\cite{Feshbach1958}.  The lattice is split into the outermost closed ring $B$ and the remaining interior sites $I$.  Only $B$ carries the boundary dissipation.  The full Hamiltonian is written as
\begin{equation}\label{eq:feshbachblock}
\begin{aligned}
&H(\gamma)=\begin{pmatrix}
H_{BB}+i\gamma G_{BB} & H_{BI}\\
H_{IB} & H_{II}
\end{pmatrix},\qquad \Psi=\begin{pmatrix}
\psi_B\\
\psi_I
\end{pmatrix}.
\end{aligned}
\end{equation}
Every block comes directly from the microscopic lattice Hamiltonian.

For a Hermitian reference doublet $\alpha$, we first define its mean energy,
\begin{equation}\label{eq:feshbachreferenceenergy}
E_\alpha^{(0)}=\frac12\operatorname{Tr}\left[Q_\alpha^{(0)\dagger}H_0Q_\alpha^{(0)}\right].
\end{equation}
The interior self-energy is evaluated once at this energy and kept fixed as the dissipation is varied:
\begin{equation}\label{eq:feshbachselfenergy}
\Sigma_\alpha^{(0)}=H_{BI}\left(E_\alpha^{(0)}-H_{II}\right)^{-1}H_{IB}.
\end{equation}
The reduced boundary Hamiltonian is then
\begin{equation}\label{eq:feshbachmain}
H_{\rm F}^{(\alpha)}(\gamma)=H_{BB}+i\gamma G_{BB}+\Sigma_\alpha^{(0)}.
\end{equation}

At $\gamma=0$, we select the reduced right eigenvectors by their overlap with the boundary part of $Q_\alpha^{(0)}$.  We reconstruct the corresponding full-lattice states as
\begin{equation}\label{eq:feshbachreconstruction}
\begin{aligned}
&\widetilde\Psi_{\rm F}^{(\alpha)}=\begin{pmatrix}
\Psi_{B}^{(\alpha)}\\
R_\alpha^{(0)}\Psi_{B}^{(\alpha)}
\end{pmatrix},\\
&R_\alpha^{(0)}=\left(E_\alpha^{(0)}-H_{II}\right)^{-1}H_{IB}.
\end{aligned}
\end{equation}

For the propagating calculation, $\alpha={\rm P}+$ denotes the positive-energy Hermitian propagating doublet.  For the corner calculation, $\alpha={\rm C}+$ and $\alpha={\rm C}-$ denote the positive- and negative-energy Hermitian corner doublets, respectively.

\end{document}